# Distinguishing coherent and thermal photon noise in a circuit QED system

Fei Yan[1][*] and Dan Campbell[1], Philip Krantz[1], Morten Kjaergaard[1],
David Kim[2], Jonilyn L. Yoder[2], David Hover[2], Adam Sears[2],
Andrew J. Kerman[2], Terry P. Orlando[1], Simon Gustavsson[1], William D. Oliver[1,2,3]

[1] Research Laboratory of Electronics, Massachusetts Institute of Technology, Cambridge, MA 02139, USA
[2] MIT Lincoln Laboratory, 244 Wood Street, Lexington, MA 02420, USA and
[3] Department of Physics, Massachusetts Institute of Technology, Cambridge, MA 02139, USA

In the cavity-QED architecture, photon number fluctuations from residual cavity photons cause qubit dephasing due to the AC Stark effect. These unwanted photons originate from a variety of sources, such as thermal radiation, leftover measurement photons, and crosstalk. Using a capacitively-shunted flux qubit coupled to a transmission line cavity, we demonstrate a method that identifies and distinguishes coherent and thermal photons based on noise-spectral reconstruction from time-domain spin-locking relaxometry. Using these measurements, we attribute the limiting dephasing source in our system to thermal photons, rather than coherent photons. By improving the cryogenic attenuation on lines leading to the cavity, we successfully suppress residual thermal photons and achieve $T_1$-limited spin-echo decay time. The spin-locking noise spectroscopy technique can readily be applied to other qubit modalities for identifying general asymmetric non-classical noise spectra.

Superconducting cavity modes are widely used in quantum information processing for mediating two-qubit interactions, reading out qubits, storing information, and even acting as qubits [1–4]. Each added cavity can also introduce additional channels for qubit relaxation and dephasing [5–7]. For example, photon number fluctuations in the cavity cause qubit dephasing due to a photon-number-dependent frequency shift of the qubit, the AC Stark effect, which is considered a major source of decoherence in state-of-the-art systems [8]. For cooling or reseting cavities [9–11], it remains challenging to identify and suppress unwanted cavity photons at the level demanded by fault-tolerant applications. These residual cavity photons arise from a variety of sources, e.g. thermal microwave photons from blackbody radiation that is improperly attenuated or filtered in the cryogenic environment [12]. Additionally, unwanted coherent photons can remain in the cavity following readout or from microwave crosstalk in a multi-qubit system [5, 13]. The distinction between thermal and coherent photons has been measured by either spectroscopically resolving photon-number states of the qubit [14, 15] or by characterizing the dependence of qubit dephasing on the mean photon number $\bar{n}$ [16] in the regime where qubit-cavity dispersive coupling $\chi$ is much stronger than the photon loss rate $\kappa$ and where the mean photon number is sufficiently large ($\bar{n} \gtrsim 1$). However, in the operating regime more relevant to quantum information processing applications, where the coupling is relatively weak ($2\chi \lesssim \kappa$) and the residual photon number is small ($\bar{n} \ll 1$), it becomes challenging to differentiate photons from uncontrollable coherent and thermal sources.

An alternative approach to characterizing and identifying a random process is the direct measurement of its power spectral densities (PSDs). There are several ways to perform noise spectroscopy [17–23], depending on the frequency range of interest and the system properties. To measure noise within a band of 0.1-100 MHz which is relevant to dephasing in superconducting qubits, the spin-locking – also called $T_{1\rho}$ – noise-spectroscopy technique was demonstrated in our previous works [8, 24] and proved to be advantageous, because it takes advantage of robustness against low-frequency noise that is inherent in driven evolution, and it features a relatively straightforward relaxometry analysis with fewer assumptions than free-evolution techniques [19]. However, the previous work was limited to classical spectra, and a general non-classical spectrum – evident as an asymmetric noise spectrum with respect to zero frequency [25] – has not yet been demonstrated. Developing such a complete characterization technique enables studies with both fundamental and practical significance, e.g., for metrology, non-equilibrium dynamics, qubit noise correlations, and environmental engineering [26]. The spin-locking protocol has also been demonstrated in coherence characterization [27], active cooling [28], and magnetic sensing [29].

In this Letter, we extend the spin-locking noise spectroscopy technique to the case of non-classical (two-sided, emission and absorption) noise spectra, and demonstrate it with engineered coherent and thermal photon noise. To the qubit, the engineered noise photons generated by a coherent source exhibit a Lorentzian-shaped spectrum centered at the frequency detuning between the cavity and the drive. Such an asymmetric spectrum is clearly traceable to the non-equilibrium, engineered-noise environment, even when the system is operated in an otherwise classical equilibrium limit ($k_B T \gg \hbar \omega$, where $T$ is temperature). The spectral width of coherent photon noise is half that of the corresponding spectral width for thermal photons, due to the difference in their correlation times. Using this technique, we find that the photon noise limiting our qubit dephasing originates from ther-mal sources. By improving the cryogenic attenuation in our experiment, we obtain a spin-echo decay time $T_{2\text{Echo}}$ near the $2T_1$ limit.

Our test device (Fig. 1a) is a superconducting capacitively shunted flux qubit (CSFQ) embedded in a coplanar transmission line cavity. Details about the design and fabrication of this device can be found in [8]. The sample is tested in a dilution refrigerator at a base temperature of 20 mK. The cavity has a center frequency $\omega_c/2\pi \approx 8.26$ GHz and decay rate $\kappa/2\pi \approx 1.69$ MHz. The qubit is biased at its flux-insensitive point (transition frequency $\omega_q/2\pi \approx 3.70$ GHz) and is capacitively coupled to the cavity, resulting in an effective dispersive coupling between the qubit and cavity (coupling strength $\chi/2\pi \approx 0.31$ MHz). An rf signal ($\omega_{\text{ro}}$) near the cavity frequency is used to read out the qubit state (Fig. 1b). Within the rotating-wave approximation [24, 28], a qubit drive ($\omega_{\text{qb}}$) creates an effective two-level system in the rotating frame with a transition frequency equivalent to the Rabi frequency $\Omega$. The Hamiltonian with resonant qubit drive ($\omega_{\text{qb}} = \omega_q$) can be expressed as

$$H/\hbar = \Omega\, \hat{\sigma}_z/2 + \chi\, \hat{a}^\dagger \hat{a}\, \hat{\sigma}_x + \omega_c\, \hat{a}^\dagger \hat{a} \,, \qquad (1)$$

where $\hat{\sigma}_z$ is the Pauli operator in the new eigenbasis defined by the Rabi drive field, and $\hat{a}^\dagger (\hat{a})$ is the cavity photon creation(annihilation) operator. The photon number operator is then $\hat{n} = \hat{a}^\dagger \hat{a}$. Eq. (1) indicates that photon number fluctuations – the same noise that dephases the original qubit in the lab frame – now transversely couple to the dressed qubit and result in energy exchange. This shows that a random process may affect a quantum system differently depending on the control protocol. Considering only lab-frame relaxation processes (relaxation rate $\Gamma_1 = 1/T_1$) and photon noise, we can relate the dressed qubit's relaxation/emission rates, $\Gamma_{\downarrow \rho}$ and $\Gamma_{\uparrow \rho}$, to the noise PSD using Fermi's Golden Rule,

$$\Gamma_{\downarrow \rho} = \chi^2\, \mathcal{S}_{nn}(\omega = \Omega) + \Gamma_1/4 \,, \qquad (2)$$

$$\Gamma_{\uparrow \rho} = \chi^2\, \mathcal{S}_{nn}(\omega = -\Omega) + \Gamma_1/4 \,, \qquad (3)$$

where $\mathcal{S}_{nn}(\omega) = \int_{-\infty}^{\infty} d\tau\, e^{-i\omega\tau} \langle \hat{n}(0)\hat{n}(\tau) \rangle$ is the photon noise PSD. For simplicity, we will omit the subscript "$nn$" throughout the rest of the manuscript. We emphasize that the fluctuating variable $\hat{n}$ is a quantum operator, and its autocorrelation function $\mathcal{C}(\tau) = \langle \hat{n}(0)\hat{n}(\tau) \rangle$ may be complex, or equivalently, its fourier transform asymmetric, $\mathcal{S}(\omega) \neq \mathcal{S}(-\omega)$. This is in contrast to classical noise, for which the above operator $\hat{n}$ is replaced with a classical variable $n$, such that $\mathcal{C}(\tau)$ is manifestly real and, thus, the power spectrum $\mathcal{S}(\omega) = \mathcal{S}(-\omega)$ is symmetric. In this work, the engineered noise from a coherent-state cavity photon serves as a non-equilibrium qubit environment, leading to an asymmetric photon noise spectrum. In contrast, for an environment in thermal equilibrium, the noise spectrum would be symmetric in the classical limit ($k_\text{B} T \gg \hbar\Omega$) and asymmetric in the quantum limit ($k_\text{B} T \ll \hbar\Omega$).

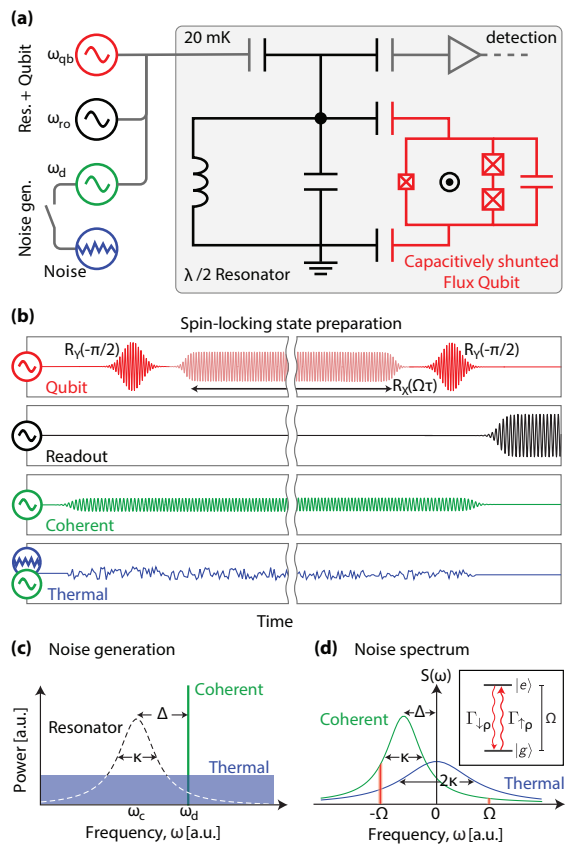

FIG. 1. **Simplified measurement schematic and noise generation.** **(a)** Device and measurement setup. A CSFQ is capacitively coupled to a $\lambda/2$ transmission line cavity. Besides qubit control (red) and readout (black) pulses, a weak, uniform, cavity drive (green) is applied to create a coherent state. To mimic thermal photon noise, we amplitude-modulate this drive with a broadband white noise (blue). **(b)** Microwave implementation of experiment. The qubit control sequence executes the spin-locking protocol, during which the photon noise generation signal is kept on. A strong readout pulse follows for dispersively measuring the qubit state. **(c)** Schematic diagram of coherent/thermal photon noise generation. The black curve indicates the cavity response. A single-frequency drive creates coherent photons in the cavity. A broadband noise (bandwidth $\gg \kappa$) mimics the thermal photon noise [30]. **(d)** Schematic of coherent/thermal photon noise spectrum. Coherent photons produce a noise spectrum symmetric with respect to $\omega = -\Delta$, and generally lead to unequal emission and absorption spectral densities (red bars), and hence unequal relaxation and excitation rates for the dressed qubit (inset). Thermal photons produce a symmetric spectrum, and the width is twice that of coherent photons.

In the experiment, we engineer both coherent and thermal photon noise environments to study their differences (Fig. 1c). Coherent photon noise is generated by sending a coherent tone at frequency $\omega_\text{d}$ ($\omega_\text{d} \sim \omega_c$) to drive and maintain a coherent state in the cavity. The corresponding photon-number correlation function is $\mathcal{C}^{\text{coh}}(\tau) = \bar{n}\, e^{-i\Delta\tau - \kappa|\tau|/2} + \bar{n}^2$, where $\Delta = \omega_d - \omega_c$ is the frequency detuning of the coherent drive from the cavity [1, 25]. Thermal photon noise is mimicked by generating a broadband white noise (bandwidth = 80 MHz) extending well beyond the cavity linewidth [14]. The corresponding photon-number correlation function is $\mathcal{C}^{\text{th}}(\tau) = (\bar{n}^2 + \bar{n})\, e^{-\kappa|\tau|} + \bar{n}^2$ [31]. In the limit of small $\bar{n}$, these two correlation functions differ only by a factor of 2 in the correlation time. For coherent photons, the dominant term in the correlation function is the fluctuating field operator contracted with a coherent field – effectively a classical field – and the resulting correlation function decays at rate $\kappa/2$. In contrast, for the thermal photon noise case, the leading term is the contraction of the fluctuating field operator with itself, and this results in a correlation function with rate $\kappa$. This characteristic difference is also reflected in their corresponding Fourier transforms (Fig. 1d),

$$\mathcal{S}^{\text{coh}}(\omega) = \bar{n}\, \frac{\kappa}{(\omega + \Delta)^2 + (\kappa/2)^2}\,, \quad (4)$$

$$\mathcal{S}^{\text{th}}(\omega) = (\bar{n}^2 + \bar{n})\, \frac{2\kappa}{\omega^2 + \kappa^2}\,, \quad (5)$$

for $\omega \neq 0$. Both spectra are Lorentzian-shaped and increase with the mean photon number (linearly if $\bar{n} \ll 1$). There are two notable differences between Eq. (4) and (5). First, the spectrum $\mathcal{S}^{\text{coh}}(\omega)$ for coherent photons is centered around $\omega = -\Delta$, indicating an asymmetric noise spectrum and, hence, emission and absorption rates that are not equal to one another in general. Second, the width of the spectral distribution with thermal photons is twice that with coherent photons (HWHM: $\kappa$ vs. $\kappa/2$), reflecting the difference in their correlation times. Note that Eqs. (4-5) are only strictly valid when the qubit dispersive shift is small relative to the cavity decay rate ($\chi \ll \kappa$). Theories about the generalized case are discussed in Ref. [13, 32].

We perform spin-locking noise spectroscopy to measure the noise PSDs in the frequency range near the cavity decay rate $\kappa$. The implemented spin-locking pulse sequence comprises three pulses (Fig. 1b). Following a $\pi/2$-pulse about the $-Y$ axis, a constant drive along the X axis locks the qubit state for a variable duration (conventional rotating-frame X/Y notation applies here). The locking X-pulse also defines the dressed quantization axis (denoted by the $z$ axis as in Eq. (1)). The longitudinal polarization $\langle \sigma_z \rangle$ decays at rate $\Gamma_{1\rho} = \Gamma_{\downarrow\rho} + \Gamma_{\uparrow\rho}$ during the locking period. A final $\pi/2$-pulse about the Y axis projects the remaining polarization onto the measurement axis. By varying the length of the constant drive, we record the decay of the $z$-polarization. Following Eqs. (2-3) and in steady-state, we have

$$\bar{\mathcal{S}}(\Omega) = \frac{\mathcal{S}(\Omega) + \mathcal{S}(-\Omega)}{2} = \frac{1}{2\chi^2}\left(\Gamma_{1\rho}(\Omega) - \frac{\Gamma_1}{2}\right)\,, \quad (6)$$

$$\langle \sigma_z^{\text{ss}}(\Omega)\rangle = \frac{\mathcal{S}(\Omega) - \mathcal{S}(-\Omega)}{\mathcal{S}(\Omega) + \mathcal{S}(-\Omega)}\,, \quad (7)$$

where $\bar{\mathcal{S}}(\omega) = \int_{-\infty}^{\infty} d\tau\, e^{-i\omega\tau} \frac{1}{2}\langle \hat{n}(0)\hat{n}(\tau) + \hat{n}(\tau)\hat{n}(0)\rangle$ is the symmetrized PSD (by definition) and $\langle \sigma_z^{\text{ss}}(\Omega)\rangle$ is the steady-state $z$-polarization. Therefore, the noise PSDs can be extracted from the $T_{1\rho}$ decay functions at experimentally feasible Rabi frequencies.

In noise measurements using a coherent driving state, we record the spin-locking decay (Fig. 2a) at various Rabi frequencies $\Omega$ and detunings $\Delta$. The traces are fit to an exponential decay, allowing us to extract $\Gamma_{1\rho}$ and $\langle \sigma_z^{\text{ss}}(\Omega)\rangle$. Using Eq. (6), the decay constant is converted to $\bar{\mathcal{S}}(\Omega, \Delta)$ and plotted in Fig. 2b. Since $\bar{\mathcal{S}}(\omega)$ is symmetric by definition, it is sufficient to show only its values at positive frequencies. The steady-state polarization for various $\Omega$ and $\Delta$ is plotted in Fig. 2c. Significant deviations from zero polarization suggests unequal relaxation and excitation rates. For the case of a blue-detuned drive ($\Delta \equiv \omega_d - \omega_c > 0$, case A in Fig. 2), the dressed qubit decays to a steady state closer to its excited state, meaning $\Gamma_{\downarrow\rho} < \Gamma_{\uparrow\rho}$ and the qubit tends to absorb energy from its environment. In contrast, the red-detuned case ($\Delta \equiv \omega_d - \omega_c < 0$, case C) saturates closer to the ground state, meaning $\Gamma_{\downarrow\rho} > \Gamma_{\uparrow\rho}$ and the qubit predominantly emits energy to its environment. This effect was demonstrated to stabilize the qubit in a pure state (here, its ground state) [28], an alternative approach to creating a non-equilibrium effective qubit temperature low enough to reach the quantum limit ($T_{\text{eff}} \ll \Omega$).

The above features are apparent in the reconstructed spectra $\mathcal{S}(\omega)$ shown in Figs. 3a and 3b, which are derived from Eqs. (6-7) at $\omega = \pm\Omega$ using the data from Fig. 2 and may in principle be non-classical. First, the smooth and continuous transition around $\omega = 0$ indicates that our spectroscopy methods correctly connects positive and negative frequencies. The tilted square pattern in the 2D spectrum $\mathcal{S}(\omega, \Delta)$ also indicates that the spectral peak moves with $\Delta$. The peak amplitude is reduced when moving away from zero frequency for the same drive power, as the mean photon number becomes smaller ($\bar{n} \propto \frac{1}{\Delta^2 + (\kappa/2)^2}$). Figs. 3c and 3d show that the extracted Lorentzian center frequencies for coherent photons follow $\omega = -\Delta$, and the HWHMs ($\approx 0.83$ MHz) are approximately half that of the corresponding measurement for thermal photons ($\approx 1.61$ MHz). Such a factor of 2 difference in spectral width is a major feature that can be used for differentiating coherent and thermal photon sources. We note that the width difference is not exactly twice, due to the small yet non-negligible perturbation from the qubit ($\chi/\kappa \approx 0.18$), which shifts the cavity response depending on the qubit state [13]. The experimental results agree with Eqs. (4-5), validating our approach.

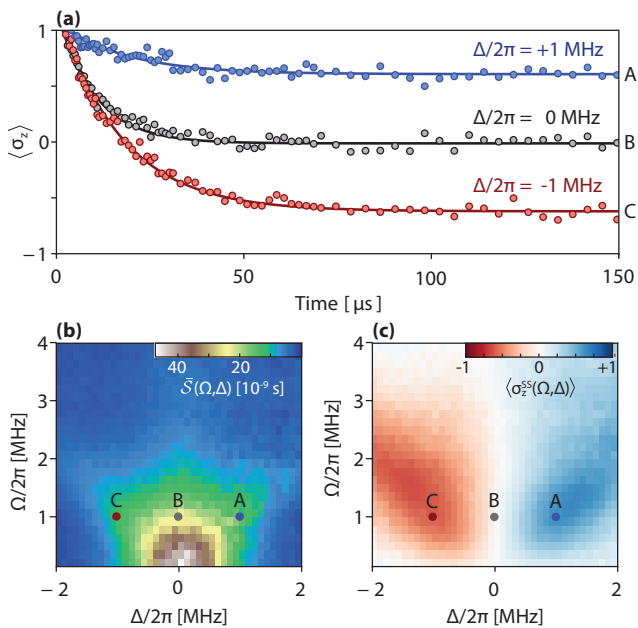

FIG. 2. $T_{1\rho}$ **decay characterization.** (a) $T_{1\rho}$ decay with coherent photons. The selected traces are measured at cavity drive detuning $\Delta/2\pi = +1, 0, -1$ MHz respectively, and with the same cavity and qubit drive power ($\Omega/2\pi = 1$ MHz). The solid lines are exponential fits. (b) Symmetrized noise PSD with coherent photons vs. $\Omega$ and $\Delta$, derived from the fitted decay constants by using Eq. (6). The markers denote the cases shown in (a). (c) Steady-state polarization with coherent photons vs. $\Omega$ and $\Delta$, extracted directly from fit. The color scale, blue/red, indicates that the dressed qubit predominantly emits/absorbs energy.

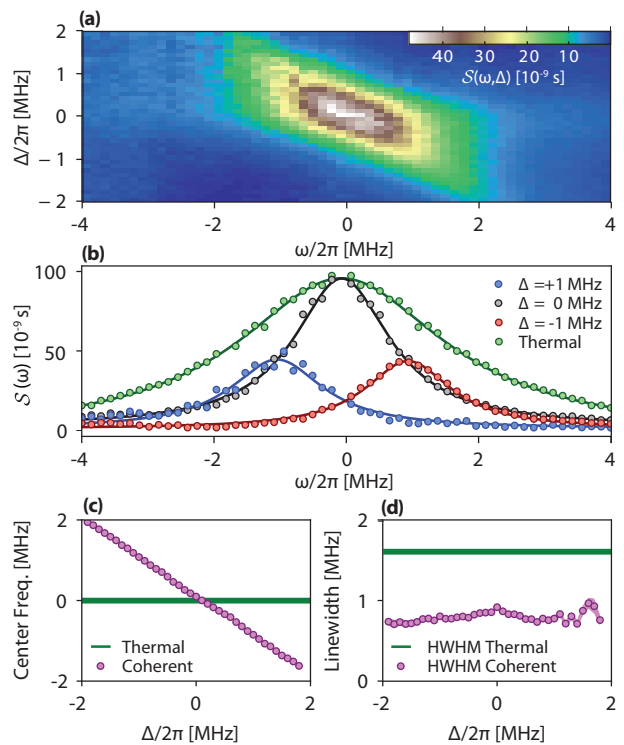

FIG. 3. **Coherent and thermal photon noise spectra.** (a) Two-sided (non-classical) PSD $\mathcal{S}(\omega, \Delta)$ with coherent photons at various $\Delta$, reconstructed from data shown in Fig. 2c and 2d. (b) Example slices of reconstructed $\mathcal{S}(\omega)$. Plotted are three typical cases with coherent drive at $\Delta/2\pi = +1, 0, -1$ MHz and one thermal case which has no $\Delta$-dependence. The solid lines are Lorentzian fit. (c) Extracted center frequencies in both coherent and thermal cases. (d) Extracted HWHMs. The mean value of HWHMs with coherent photons is 0.83 MHz. The HWHM with thermal photons is 1.61 MHz. The green lines in (c) and (d) are included as a guide to the eye, since the thermal photons are generated with broadband noise and have no $\Delta$ dependence.

We implemented the same technique on a second device with a similar design. The device was measured with $\omega_c/2\pi = 8.27$ GHz, $\kappa/2\pi = 1.5$ MHz, $\chi/2\pi = 0.45$ MHz, $\omega_q/2\pi = 4.7$ GHz, $T_1 = 35-55\,\mu$s and $T_{2\text{Echo}} = 40\,\mu$s [8]. Fig. 4a shows the spin-locking noise spectroscopy results, which again demonstrate the characteristic factor of 2 difference between the HWHMs of injected coherent and thermal photons. We also found that the spectrum measured without added noise (blue) has a -3 dB point consistent with thermal photon noise. Therefore, we suspected that thermal radiation from higher-temperature stages in the DR were responsible for the residual cavity photons and the resulting dephasing. By measuring the dependence of the dephasing rate on the average number of engineered thermal-noise photons, we extrapolated the average residual thermal photon number in the absence of externally applied noise to be around 0.006, corresponding to an 80 mK equilibrium temperature [8]. During a subsequent thermal-cycling of the dilution fridge, we increased the attenuation at the mixing chamber (20 mK) from 23 dB to 43 dB in order to reduce the thermal photons reaching the cavity (see details in supplement [34]). This modification significantly increased $T_{2\text{Echo}}$ to 80 $\mu$s (Fig. 4b), while $T_1$ did not change. The new attenuator configuration effectively suppresses the residual thermal photons in our cavity to $\bar{n} < 0.0006$ [34], ten times lower than the previous level. This corresponds to an equivalent equilibrium temperature of 55 mK. Due to temporal spread of coherence times and measurement uncertainty, we could not confirm a lower bound, though measurements of the excited-state population of several qubits tested in the same dilution fridge found an effective temperature of 35 mK [35].

To conclude, we developed a spin-locking ($T_{1\rho}$) technique for performing non-classical noise spectroscopy and demonstrated it using engineered photon noise applied to a superconducting circuit QED system. The measured noise spectra were used to distinguish between coherent and thermal cavity photon noise. Using this technique, we found that residual thermal photons limit our qubit coherence. By optimizing cryogenic attention chain, we successfully reduced the residual thermal photon number and recovered $T_{2\mathrm{Echo}} \approx 2T_1$. Further improvement may be possible with more careful choice of attenuator/filter configuration or active cooling techniques.

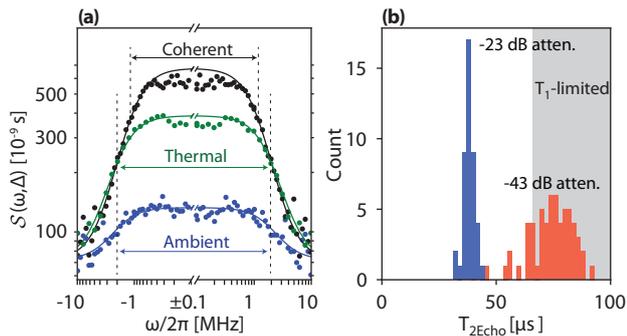

FIG. 4. **Decoherence mitigation.** (a) Photon noise spectra with the second device. Plotted are spectra with injected coherent photons (black), injected thermal photons (green), and no injected photons (blue). Solid lines are Lorentzian fit. Arrows indicate the peak width spanned by the corresponding -3 dB points (dashed lines) for the thermal and ambient cases (HWHM ≈ 2.0 MHz), and for the coherent case (HWHM ≈ 1.1 MHz). Measured PSDs at the low frequency end ($< 0.4$ MHz) are consistently lower than fit, because the injected noise disrupts the spin-locking condition when the locking drive is weak. (b) Histogram of measured $T_{2\mathrm{Echo}}$ statistics before and after adding extra attenuation on the input side of the cavity. In both cases, the fluctuation mainly comes from temporal $T_1$ instability due to quasiparticles [33]. Grey area indicates the spread of the $2T_1$ limit measured with -43 dB attenuation.

We thank Alexandre Blais, Lorenza Viola, Jan Goetz, Michael Marthaler for insightful discussions. We thank Mirabella Pulido for generous assistance. This research was funded in part by the Office of the Director of National Intelligence (ODNI), Intelligence Advanced Research Projects Activity (IARPA) and by the Assistant Secretary of Defense for Research & Engineering via MIT Lincoln Laboratory under Air Force Contract No. FA8721-05-C-0002; by the U.S. Army Research Office Grant No. W911NF-14-1-0682; and by the National Science Foundation Grant No. PHY-1415514. M.K. gratefully acknowledges support from the Carlsberg Foundation. The views and conclusions contained herein are those of the authors and should not be interpreted as necessarily representing the official policies or endorsements, either expressed or implied, of ODNI, IARPA, or the US Government.

F.Y. and D.C. contributed equally to this work.

* fyan@mit.edu

# Supplementary Materials:

# Distinguishing coherent and thermal photon noise in a circuit QED system

Fei Yan[1], Dan Campbell[1], Philip Krantz[1], Morten Kjaergaard[1],
David Kim[2], Jonilyn L. Yoder[2], David Hover[2], Adam Sears[2]
Andrew J. Kerman[2], Terry P. Orlando[1], Simon Gustavsson[1], William D. Oliver[1,2,3]

[1]*Research Laboratory of Electronics, Massachusetts Institute of Technology, Cambridge, MA 02139, USA*
[2]*MIT Lincoln Laboratory, 244 Wood Street, Lexington, MA 02420, USA and*
[3]*Department of Physics, Massachusetts Institute of Technology, Cambridge, MA 02139, USA*

## A. OPTIMIZING CRYOGENIC ATTENUATION

### A1. Analysis of blackbody radiation in a fridge

Dissipative components such as attenuators can generate blackbody radiation. To suppress qubit dephasing due to thermal photons in the cavity, one needs to reduce the overall amount of radiation that reaches the device from various temperature stages in a DR. The Johnson-Nyquist noise from a resistor $R$ at its equilibrium temperature $T$ has a voltage-noise PSD, which takes a compact form,

$$\mathcal{S}_{VV}(f; R, T) = 4k_\mathrm{B} T R \frac{hf/k_\mathrm{B}T}{e^{hf/k_\mathrm{B}T} - 1} \ . \tag{S1}$$

The form spans over both the classical (low-frequency, high-temperature) and quantum limits (high-frequency, low-temperature). Note that for $f > 0$, $\mathcal{S}_{VV}(f)$ corresponds to the emission spectrum of the resistor. After accounting for attenuation along the line, $\mathcal{S}_{VV}(f)$ also represents the absorption spectrum felt by the cavity. The noise power determines the number of residual cavity photons. The overall contribution to a cavity (frequency $f_\mathrm{c} = 8.2\,\mathrm{GHz}$) from thermal sources ($R^{(i)} = 50\,\Omega$) at different temperature stages $T^{(i)}$ is then

$$\mathcal{S}_{VV}(f_\mathrm{c}) = \sum_k A^{(i)} \mathcal{S}_{VV}(f_\mathrm{c}; R^{(i)}, T^{(i)}) \ , \tag{S2}$$

where $A^{(i)}$ accounts for the collective attenuation after the corresponding source and before reaching the cavity.

The original setup for measuring the second device (described in the main text) is depicted in in Fig. S1a. The radiation sources are located at $300\,\mathrm{K}$, $3\,\mathrm{K}$, $0.8\,\mathrm{K}$ and $20\,\mathrm{mK}$. The corresponding collective attenuation (including cable loss) are $58\,\mathrm{dB}$, $35\,\mathrm{dB}$, $26\,\mathrm{dB}$ and $0\,\mathrm{dB}$ respectively. Using Eqs. (S1-S2), we derive the effective individual PSDs (Fig. S2a) and their summation (Fig. S2b). Apparently, the most relevant contributions at the cavity frequency are from $0.8\,\mathrm{K}$, $3\,\mathrm{K}$ and $300\,\mathrm{K}$, rather than $20\,\mathrm{mK}$, resulting in a much higher effective noise temperature ($\approx 80\,\mathrm{mK}$, equivalent to an average photon number of 0.006). This explains the observed residual dephasing that restricted $T_\mathrm{2Echo}$ from the reaching $2T_1$ limit as discussed in the main text.

During the subsequent thermal-cycling, we modified the setup (Fig. S1b). In particular, we added extra $20\,\mathrm{dB}$ attenuation on the input port of the cavity at $20\,\mathrm{mK}$. This attenuator effectively reduces all of the radiation from above the mixing chamber by a factor of 100 (Fig. S2c), resulting in a lower effective temperature ($<40\,\mathrm{mK}$).

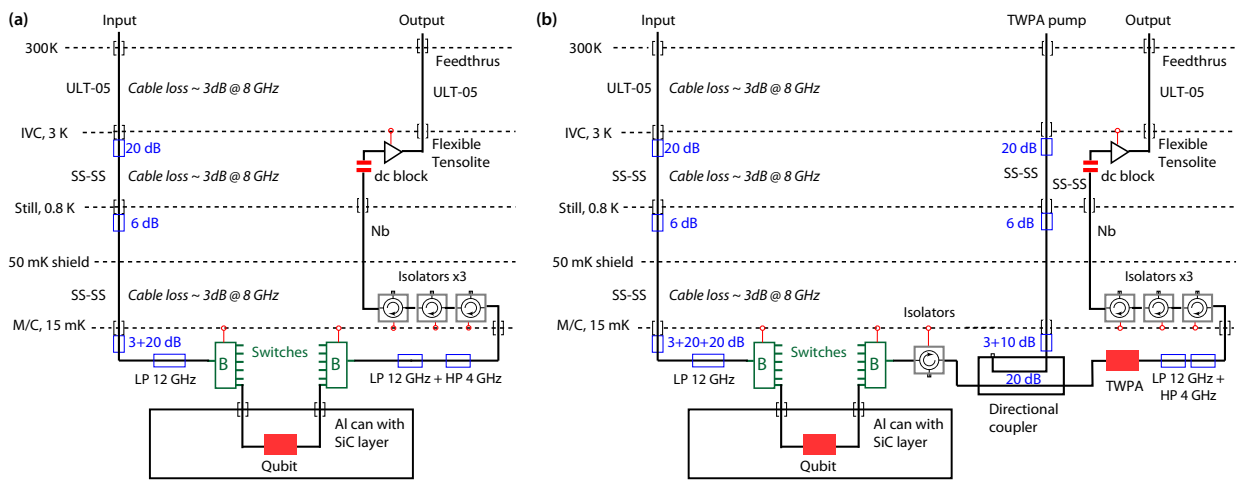

FIG. S1. **Schematics of the measurement chain inside DR before (a) and after (b) changes made to improve coherence of the second device.** The most relevant modification is an extra $20\,\mathrm{dB}$ attenuator on the line leading to the input port of the cavity at the mixing chamber (15-20 mK). Following the output port of the cavity include other changes: the addition of an isolator, a directional coupler, and a traveling-wave parametric amplifier (TWPA).

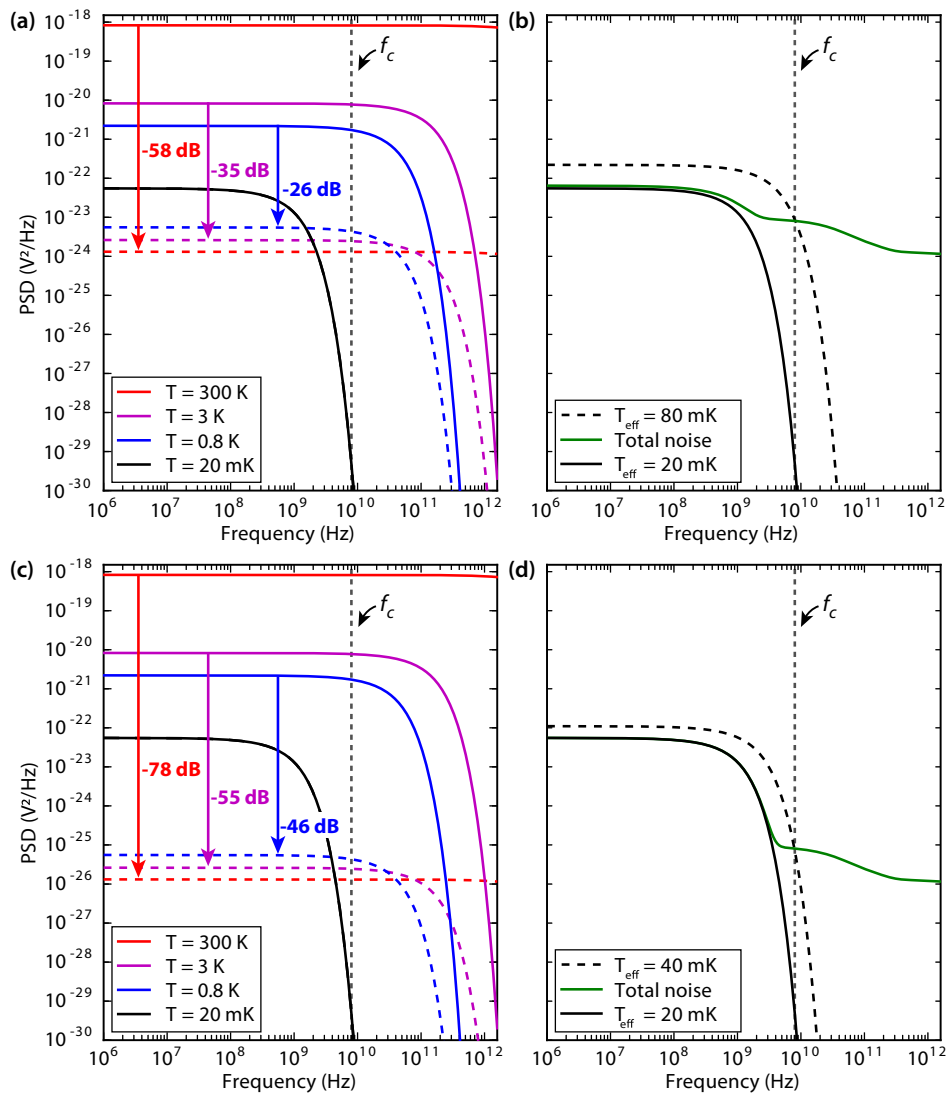

FIG. S2. **Noise PSDs of thermal radiation from different temperature stages.** **a** Individual contributions before change of setup. The noise PSDs generated at the sources (solid lines) follow Eq. S1. Their equivalent PSDs at the mixing chamber (dashed lines) are calculated after taking into account attenuation from all the attenuators and cables before reaching the cavity. The black dash line indicates the cavity frequency. **b** Total contributions at the mixing chamber before change of setup . The case of an 80 mK source is plotted for comparison. **c** Individual contributions after adding 20 dB attenuation. **d** Total contribution at the mixing chamber after adding 20 dB attenuation. The case of a 40 mK source is plotted for comparison.

### A2. Residual photon number estimate from dephasing

We characterized the same qubit after the change of configuration. Fig. S3 shows the relaxation, echo, and derived pure dephasing rates corresponding to the data shown in Fig. 4b of the main text as a function of repetition number. The data is taken with interleaved $T_1$ and spin echo measurements, so as to remove the effect from temporal $T_1$ fluctuation. We note there are a few outliers having abnormally high echo decay rates. Including these outliers, the derived pure dephasing rates have a mean value $1.6 \times 10^3 \mathrm{s}^{-1}$ and standard deviation $2.7 \times 10^3 \mathrm{s}^{-1}$. The mean value approximately corresponds to an average residual photon number of 0.0006 and an equivalent temperature of 55 mK. Since the statistical uncertainty is greater than the mean value, we could not confirm any lower number by this method.

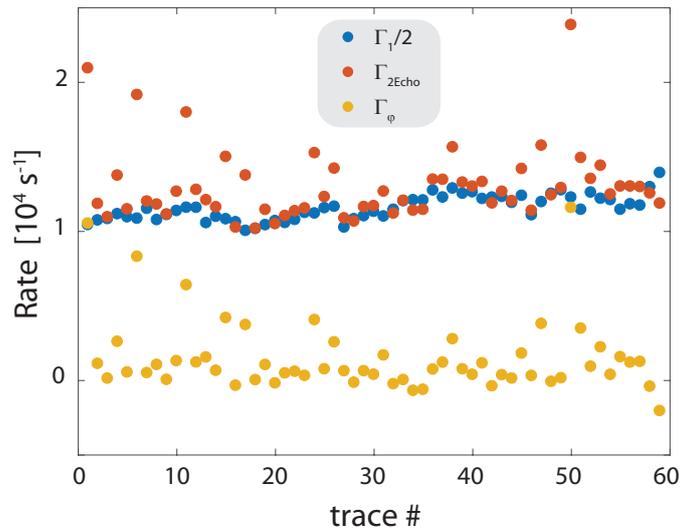

FIG. S3. Relaxation ($\Gamma_1$) and spin-echo ($\Gamma_{2\text{Echo}}$) decay rates (exponential) extracted from repeatedly interleaved measurements. Pure dephasing rates ($\Gamma_\phi$) are subsequently derived from $\Gamma_\varphi = \Gamma_{2\text{Echo}} - \frac{1}{2}\Gamma_1$. The measurement takes about six hours.

## B. CORRELATION OF CAVITY PHOTON

In this section, we study the two-time correlation function of the photon number in a cavity,

$$\begin{aligned}\mathcal{C}(\tau) &= \langle \hat{n}(t+\tau)\hat{n}(t)\rangle \\ &= \langle \hat{a}^\dagger(t+\tau)\hat{a}(t+\tau)\hat{a}^\dagger(t)\hat{a}(t)\rangle\,,\end{aligned} \quad (S3)$$

where the field operator $\hat{a}(t)$ and $a^\dagger(t)$ are defined with respect to the cavitys rotating frame. $\hat{n}(t) = \hat{a}^\dagger(t)\hat{a}(t)$ is the associated photon number operator. This correlation function may be alternatively represented by its Fourier transform, the power spectrum of photon number fluctuations $\mathcal{S}(\omega) = \int_{-\infty}^{\infty} d\tau e^{-i\omega\tau}\mathcal{C}(\tau)$. We are interested in the statistical description of the noise injected into a cavity from thermal and coherent sources.

### B1. Stationary noise at the input of the cavity

Making use of the generator of quantization noise in the cavity $\left[\hat{a}(t+\tau), \hat{a}^\dagger(t)\right] = e^{-(\kappa/2)|\tau|}$ (see derivations in section B2), we may rewrite Eq. S3 as:

$$\mathcal{C}(\tau) = \langle \hat{a}^\dagger(t)\hat{a}^\dagger(t+\tau)\hat{a}(t+\tau)\hat{a}(t)\rangle + \langle \hat{a}^\dagger(t+\tau)\hat{a}(t)\rangle e^{-(\kappa/2)|\tau|}\,. \quad (S4)$$

The first term on the RHS is sometimes called the wave contribution. It is analogous to the variance of the classical photon number when $\langle \hat{a}\rangle = 0$. The second term is sometimes called the particle contribution, which is analogous to classical photon shot noise [2].

It is convenient to separate the cavity field mode into photonic excitation and vacuum fluctuation contributions:

$$\hat{a}(t) = \bar{a}(t) + \hat{d}(t)\,. \quad (S5)$$

This approach is valid when the excitation amplitude $\bar{a}(t)$ and the quantization noise are uncorrelated; such is the case for unsqueezed coherent and thermal excitations. The emergent differences between the statistics and time dependence of coherent and thermal excitations therefore lie entirely with the time dependence of $\bar{a}(t)$ since $\hat{d}(t)$ is common to both. The amplitude of a coherent excitation is time-independent, whereas the amplitude of a thermal excitation convolves the broadband fluctuations outside the cavity with the cavity's linewidth. Substituting Eq. (S5) into Eq. (S4), we have

$$\mathcal{C}(\tau) = \bar{a}^*(t)\bar{a}^*(t+\tau)\bar{a}(t+\tau)\bar{a}(t) + \bar{a}^*(t+\tau)\bar{a}(t)e^{-(\kappa/2)|\tau|}, \qquad (S6)$$

where we eliminate the terms that annihilate the vacuum: $\langle\hat{d}(t)\rangle = \langle\hat{d}^\dagger(t)\rangle = 0$; and $\langle\hat{d}^\dagger(t+\tau)\hat{d}(t)\rangle = 0$.

In the case of coherent driving that is detuned from the cavity frequency by $\Delta = \omega_d - \omega_c$, we obtain the relevant field-field correlations:

$$\text{coherent:} \quad \langle\hat{a}^\dagger(t+\tau)\hat{a}(t)\rangle = \bar{a}^*(t+\tau)\bar{a}(t) = \bar{n}\,e^{-i\Delta\tau}. \qquad (S7)$$

The field-field correlation part of the particle contribution term is time-independent for coherent fields, the correlation function simplifies to $\mathcal{C}^{\text{coh}}(\tau) = \bar{n}\,e^{-(\kappa/2)|\tau|} + \bar{n}^2$, where the only contribution to time-dependence is from the quantization noise.

By contrast, for stationary thermal noise or chaotic light, the noise source is much more broadband than the cavity linewidth. Therefore, the cavity field-field fluctuations are capped by the cavity amplitude relaxation rate $\kappa/2$, that is,

$$\text{thermal:} \quad \langle\hat{a}^\dagger(t+\tau)\hat{a}(t)\rangle = \bar{a}^*(t+\tau)\bar{a}(t) = \bar{n}\,e^{-(\kappa/2)|\tau|}. \qquad (S8)$$

Plugging into Eq. (S6), the wave contribution takes the square of the field-field correlations, while the particle contribution takes the product of quantization noise with time varying field-field correlations. Both contributions have a correlation damping rate $\kappa$, twice that of the coherent case. The correlation function simplifies to $\mathcal{C}^{\text{th}}(\tau) = (\bar{n}^2 + \bar{n})\,e^{-\kappa|\tau|} + \bar{n}^2$.

### B2. Chaotic light and thermal photons

In our setup, we modulate the envelope of a coherent microwave tone near cavity frequency ($\sim 8.26\,\text{GHz}$) with 80 MHz broadband Gaussian noise to produce chaotic light, which is a substitute for thermal radiation. By contrast with thermal radiation, which is spectrally broad and obeys the Planck distribution, a chaotic light will appear coherent if it is measured on a sufficiently short timescale, e.g., $t \ll \tau_{\text{coh}} = 1/(2\pi \times 80\,\text{MHz})$. We examine the statistics of the cavity mode in the presence of stationary driving by chaotic light in the limit of $t \approx 1/\kappa \gg \tau_{\text{coh}}$ and compare the result to that of thermal equilibrium.

We initially consider the driving field $\hat{b}_{in}$ in the continuum of modes outside the cavity. For convenience, we define these operators such that the instantaneous power is $P_d(t) = \hbar\omega_d\,\hat{b}_{in}^\dagger(t)\hat{b}_{in}(t)$ and the commutation relation is $\left[\hat{b}_{in}(t), \hat{b}_{in}^\dagger(t)\right] = \left[\hat{b}_{out}(t), \hat{b}_{out}^\dagger(t)\right] = \kappa\left[\hat{a}(t), \hat{a}^\dagger(t)\right]$. We write down the time evolution of the cavity mode in the rotating frame [1]:

$$\hat{a}(t) = e^{-(\kappa/2)(t-t_0)}\hat{a}(t_0) + \sqrt{\kappa}\int_{t_0}^{t} dt'\,e^{-(\kappa/2)(t-t')}\hat{b}(t'). \qquad (S9)$$

The first term corresponds to the decay of the cavity mode starting at $t_0$; the second term corresponds to a convolution of the driving field with a filter function $e^{-(\kappa/2)\delta t}\Theta(\delta t)$, where $\delta t = t - t'$. The presence of this filter function is more obvious in the frequency domain and in the limit where $t_0 \to -\infty$. Applying the convolution theorem to the second term on the RHS of Eq. (S9), we obtain for $\hat{a}(t) = \bar{a}(t) + \hat{d}(t)$

$$\bar{a}(\Delta) = \mathcal{F}[e^{-(\kappa/2)(\delta t)}\Theta(\delta t)]\,\bar{b}(\Delta) = \frac{\sqrt{\kappa}}{(\kappa/2) + i\Delta}\bar{b}(\Delta), \qquad (S10)$$

$$\hat{d}(\Delta) = \mathcal{F}[e^{-(\kappa/2)(\delta t)}\Theta(\delta t)]\,\hat{\xi} = \frac{\sqrt{\kappa}}{(\kappa/2) + i\Delta}\hat{\xi}, \qquad (S11)$$

where the Lorentzian filter is applied to all excitations at the input of the cavity and the generator of vacuum fluctuations in the waveguide is given by $\left[\hat{\xi}(t+\tau), \hat{\xi}(t)\right] = \delta(\tau)$.

In a waveguide, we may distinguish chaotic light from thermal light by its relatively narrow spectral width. However, this distinction is lost with respect to the cavity mode, because the Lorentzian filter is only sensitive to excitations in a narrow band about the cavity frequency. Chaotic light and thermal excitations are indistinguishable so long as $\bar{b}(\Delta)$ is flat about the cavity frequency.

In the opposite case, when $\bar{b}(\Delta)$ is narrower than the cavity linewidth, that is, a coherent excitation, $\mathcal{F}[e^{-(\kappa/2)\delta t}\Theta(\delta t)]\hat{\xi}$ becomes the sole contributor to fluctuations in the cavity field. In this context, the time dependence in the correlation function is due to the vacuum noise of the cavity. Since it appears in the particle contribution of the photon noise spectrum, it is sometimes called the quantization noise.

### B3. Instantaneous correlation after turning off the coherent drive

Our technique for spectrally distinguishing coherent and thermal excitations in a cavity applies to a steady-state environment. In this part, we show that when we break the steady-state assumptions by suddenly turning off the coherent drive to the cavity, the instantaneous spectral width will then have a HWHM of $\kappa$, rather than the steady-state value of $\kappa/2$ before turning-off.

Using equation (S9), we evaluate the case where the coherent drive on a cavity is suddenly turned off at time $t_0$,

$$\hat{a}(t) = e^{-(\kappa/2)(t-t_0)}\bar{a}(t_0) + \hat{d}(t_0) + \sqrt{\kappa}\int_{t_0}^{t} dt' e^{-(\kappa/2)(t-t')}\hat{\xi}(t'). \tag{S12}$$

Note that before turning off the drive, $\bar{a}(t)$ is at its steady-state value. Right after turning-off, $\bar{a}(t)$ decays exponentially and contributes to the time dependence of the field-field correlation, $\langle \hat{a}^\dagger(t_0+\tau)\hat{a}(t_0)\rangle = e^{i\Delta\tau}(\Theta(-\tau)+\Theta(\tau)e^{-(\kappa/2)\tau})$, in both the particle and wave parts of the two-time correlation function. Eq. (S6) becomes

$$\begin{aligned}\mathcal{C}^{\text{coh}}(\tau) &= \bar{n}(t_0+\tau)\bar{n}(t_0) + \bar{a}^*(t_0+\tau)\bar{a}(t_0)e^{-(\kappa/2)|\tau|} \\ &= \bar{n}^2(t_0)(\Theta(-\tau)+\Theta(\tau)e^{-\kappa\tau}) + \bar{n}(t_0)e^{i\Delta\tau-(\kappa/2)|\tau|}(\Theta(-\tau)+\Theta(\tau)e^{-(\kappa/2)\tau}).\end{aligned} \tag{S13}$$

If the measurement of correlation is conducted after turning off the drive, the correlation decay rate or the spectral width is indistinguishable from that of broadband chaotic light or thermal excitations.

### B4. Commutation relation for quantization noise in a cavity.

In this part, we prove the commutation relation of quantization noise in the cavity, $[\hat{a}(t+\tau), \hat{a}^\dagger(t)] = e^{-(\kappa/2)|\tau|}$, where $\kappa/2$ is the cavity's amplitude relaxation rate. The expression is inspired by a derivation in Ref. [1]. For the case of white vacuum noise, the vacuum field operator $\hat{\xi}(t)$ has no correlation between different times. The quantum commutation relation is then

$$\left[\hat{\xi}(t+\tau), \hat{\xi}^\dagger(t)\right] = \delta(\tau). \tag{S14}$$

The relationship between the vacuum field operators in the cavity and the field operators in the waveguide is given by Eq. (S9) in the $t_0 \to \infty$ limit

$$\hat{d}(t) = \sqrt{\kappa}\int_{-\infty}^{t} dt' e^{-(\kappa/2)(t-t')}\hat{\xi}(t'). \tag{S15}$$

By direct substitution we solve for the two-time correlation function for the cavity vacuum field operators to obtain our expression for the generator of quantization noise in the cavity:

$$\left[\hat{d}(t+\tau), \hat{d}^\dagger(t)\right] = \kappa\int_{-\infty}^{t+\tau} dt' \int_{-\infty}^{t} dt'' e^{-(\kappa/2)(t+\tau-t')}e^{-(\kappa/2)(t-t'')}\left[\hat{\xi}(t'), \hat{\xi}^\dagger(t'')\right] = e^{-(\kappa/2)|\tau|}. \tag{S16}$$

If the photon field excitations are statistically independent of the vacuum field fluctuations it is permissible to make the transformation $\hat{a} = \bar{a} + \hat{d}$. This allows a more general form of the commutation relation: $[\hat{a}(t+\tau), \hat{a}^\dagger(t)] = e^{-(\kappa/2)|\tau|}$.

---